\documentclass[11pt]{article}
\usepackage{amsmath,amssymb}

\begin{document}
\tolerance=5000
\def\bt{\beta}
\def\al{\alpha}
\def\pl{\partial}
\def\gm{\gamma}
\def\dl{\delta}
\def\th{\theta}
\def\ep{\epsilon}
\def\ad{\al\bt\gm\dl}
\newcommand{\s}{\hspace{0.15cm}}
\newcommand{\bc}{\bigcirc}

\begin{titlepage}
\begin{flushright}
    NSF-KITP-05-68   \\
    hep-th/0508204  \\
    August 2005
\end{flushright}
\begin{center}
  \vspace{3cm}
  {\bf \Large $R^4$ Corrections to Eleven Dimensional Supergravity \\
   via Supersymmetry}
  \\  \vspace{2cm}
  Yoshifumi Hyakutake\footnote{E-mail: hyaku@kitp.ucsb.edu}
  and Sachiko Ogushi\footnote{E-mail: ogushi@kitp.ucsb.edu}
   \\ \vspace{1cm}
   {\it Kavli Institute for Theoretical Physics, \\
   University of California, Santa Barbara, CA 93106, USA}
\end{center}

\vspace{2cm}
\begin{abstract}
  
By applying Noether method to $\mathcal{N}=1$ 
local supersymmetry in eleven dimensions,  we obtained two candidates 
of $R^4$ corrections to the supergravity. The bosonic parts of these two 
completely match with the results obtained by type IIA string perturbative calculations.
We also obtained 13 parameters which relate only fermionic terms.

\end{abstract}
\end{titlepage}

\setlength{\baselineskip}{0.65cm}

\section{Introduction}

M-theory is a theory of membrane with $\mathcal{N}=1$ supersymmetry in eleven dimensions,
and the low energy effective action is described by $\mathcal{N}=1$, $D=11$ supergravity.
Since M-theory is thought to be a strong coupling limit of type IIA superstring theory,
many efforts have been made to formulate it as a theory of membrane. Although we do not yet know 
the perturbative formulation of M-theory, we expect that this theory contains higher derivative
corrections to the $\mathcal{N}=1$, $D=11$ supergravity. 

In fact these higher derivative corrections are expected from the perturbative analyses
of type IIA superstring theory. According to these analyses a part of corrections to 
type IIA supergravity is given by
\begin{alignat}{3}
  &\mathcal{L}_{(\alpha')^3} \sim e^{-2\phi} I_{\text{tree}} + c \, I_{\text{1-loop}}, \label{eq:IIA}
  \\
  &I_{\text{tree}} = t_8 t_8 e R^4 + \tfrac{1}{4\cdot 2!} \epsilon_{10}\epsilon_{10} e R^4, \notag
  \\
  &I_{\text{1-loop}} = t_8 t_8 e R^4 - \tfrac{1}{4\cdot 2!} \epsilon_{10}\epsilon_{10} e R^4 
  - \tfrac{1}{6} \epsilon_{10}t_8 BR^4, \notag
\end{alignat}
where $c$ is some constant. The tree level effective action is obtained by
the four graviton amplitude and the sigma-model computation\cite{GW,GSl,GVZ}. The first two terms of 
the one-loop effective action is found by the four graviton amplitude\cite{KP}.
The last term in the one-loop effective action is introduced to ensure the string-string 
duality between type IIA on K3 and heterotic string on $T^4$\cite{VW,DLM}.
Under this duality, the last term is related to the Green-Schwarz anomaly cancellation term
in the heterotic string effective action\cite{GS}.

Thus we obtain a part of the higher derivative corrections to $\mathcal{N}=1$, $D=11$ supergravity
by lifting the result~(\ref{eq:IIA}) to eleven dimensions. Then there are two candidates which 
will be invariant under the local supersymmetry\cite{KP,Ts}(See eq.~(\ref{eq:a}) and eq.~(\ref{eq:b}))
\begin{alignat}{3}
  t_8 t_8 e R^4 + \tfrac{1}{4!} \epsilon_{11}\epsilon_{11} e R^4, \qquad
  t_8 t_8 e R^4 - \tfrac{1}{12} \epsilon_{11}t_8 AR^4. \label{eq:sinv}
\end{alignat}
Actually fermionic counter terms for the latter is obtained by employing Noether method in 
refs.~\cite{RSW1}-\cite{PVW}.
But we do not yet know the fermionic part of the former superinvariant.
In this paper we extend the efforts in refs.~\cite{RSW1}-\cite{PVW2} and 
show that the above two terms are really
obtained by the local supersymmetry. Since our ansatz for the action includes more than
one hundred terms, we employed a computer program to check the cancellation of variations.

The contents of our paper is as follows. In section 2 we briefly review the cancellation mechanism
of $\mathcal{N}=1$, $D=11$ supergravity and fix our conventions and notations.
In section 3 the ansatz for the higher derivative corrections is explained.
Especially we focus on the ansatz for bosonic terms, since this part is different from that in 
ref.~\cite{RSW2}.
The cancellation mechanism for the variations is also sketched.
In section 4 we show the results obtained by employing the computer programming.
We find only two solutions which relate bosonic terms and fermionic terms.
It is expected but still surprising that the bosonic parts completely match with the above statement.
Section 5 is devoted to conclusions and discussions.

\section{Brief Review of $\mathcal{N}=1$, $D=11$ Supergravity}

M-theory possesses $\mathcal{N}=1$ supersymmetry in 11 dimensions and its low energy
effective action is described by $\mathcal{N}=1$, $D=11$ supergravity.
In this section we briefly review the local supersymmetry transformations of this supergravity
theory while fixing our conventions and notations. 

$\mathcal{N}=1$, $d=11$ supergravity consists of three massless fields: the vielbein ${e^\mu}_i$, 
a three form potential $A_{\mu\nu\rho}$ and a Majorana gravitino $\psi_{\mu}$\cite{CJS}.
Here Greek indices are coordinates for space-time and Latin indices are those for local Lorentz. 
Both of them run from $0$ to $10$.
In this paper we consider the Lagrangian up to the order of $\mathcal{O}(\psi^4)$,
since we focus on the cancellation of the variations of bosonic terms.
Then the Lagrangian of the supergravity is given by
\begin{alignat}{3}
  \mathcal{L} = & + \mathcal{L}[eR] && + \mathcal{L}[e\bar{\psi}\psi_{(2)}] + \mathcal{L}[eF^2]
  + \mathcal{L}[eF\bar{\psi}\psi] + \mathcal{L}[AF^2] + \mathcal{O}(\psi^4), 
  \\[0.2cm]
  &\mathcal{L}[eR] &&= + eR, \notag
  \\
  &\mathcal{L}[e\bar{\psi}\psi_{(2)}] &&= - \tfrac{1}{2} e \bar{\psi}_{\rho} 
  \gamma^{\rho \mu \nu} \psi_{\mu\nu}, \notag
  \\
  &\mathcal{L}[eF^2] &&= -\tfrac{1}{2\cdot 4!} e F_{\mu\nu\rho\sigma}F^{\mu\nu\rho\sigma}, \notag
  \\
  &\mathcal{L}[eF\bar{\psi}\psi] &&= - \tfrac{1}{2\cdot2\cdot 4!} e  F_{\al\bt\gm\sigma}
  (\bar{\psi}_{\mu} \gamma^{\mu\nu\al\bt\gm\sigma} \psi_{\nu} 
  + 12 \bar{\psi}^{[\alpha} \gamma^{\beta\gamma} \psi^{\delta]} ), \notag
  \\
  &\mathcal{L}[AF^2] &&= - \tfrac{1}{6\cdot 3!(4!)^2} \epsilon_{11}^{\mu_1 \cdots \mu_{11}}
  A_{\mu_1\mu_2\mu_3} F_{\mu_4\cdots\mu_7} F_{\mu_8\cdots\mu_{11}} , \notag
\end{alignat}
where $F_{\mu\nu\rho\sigma} = 4 \partial_{[\mu}A_{\nu\rho\sigma]}$ and 
$\psi_{\mu\nu}=2D_{[\mu} \psi_{\nu]}=D_\mu \psi_\nu - D_\nu \psi_\mu$. 
The covariant derivative on the Majorana gravitino
is defined as $D_{\mu} = \partial_{\mu} + \tfrac{1}{4}\omega_{\mu}{}^{ab}\gamma_{ab}$. 
And the notation $[\cdots]$ is used to abbreviate the gamma matrices and indices in the bracket.
$\psi_{(2)}$ is the abbreviation of $\psi_{\mu\nu}$.

Now we introduce a space-time dependent parameter $\epsilon$ which transforms as a Majorana spinor.
By using this parameter the supersymmetry transformations of the massless fields are expressed as
\begin{alignat}{3}
  \delta e^{\mu}{}_i &= -\bar{\ep} \gamma^{\mu} \psi_i, \notag
  \\
  \delta \psi_{\mu} &= 2D_{\mu} \ep + \tfrac{1}{144} F_{\al\bt\gm\sigma}
  (\gamma^{\al\bt\gm\sigma}{}{}{}{}_{\mu} - 8 \delta^{\al}_{\mu} \gamma^{\bt\gm\sigma} ) \epsilon
  + \mathcal{O}(\psi^2),
  \\
  \delta A_{\mu\nu\rho} &= 
  - 3 \bar{\epsilon}\gamma_{[\mu\nu}\psi_{\rho]}. \notag
\end{alignat}
Again terms of the order of $\mathcal{O}(\psi^2)$ are neglected, 
since we are considering the cancellation of the variations which are linear to $\psi$.
In order to check the supersymmetry for the supergravity, we employ the 1.5 order formalism.
So the variations of the spin connection are trivially cancelled after we write it in terms of 
the vielbein and the gravitino, i.e., $\omega_\mu{}^{ab}(e,\psi)$.
However, when higher derivative corrections are added to the supergravity,
the variations of the spin connection do not cancel automatically.
And we need the supersymmetry transformation for supercovariant spin connection of the form
\begin{alignat}{3}
  \delta \hat{\omega}_\mu{}^{ab} &= - \tfrac{1}{2} \bar{\epsilon} \gamma_\mu \psi^{ab}
  + \tfrac{1}{2} \bar{\epsilon} \gamma^a \psi^b{}_{\mu}
  - \tfrac{1}{2} \bar{\epsilon} \gamma^b \psi^a{}_{\mu}.
\end{alignat}
The symbol $\hat{\omega}_\mu{}^{ab}$ represents the supercovariant spin connection, whose 
supersymmetry variation does not include derivatives of $\epsilon$.
The Riemann tensors in the higher derivative terms are defined by using this supercovariant
spin connection.

The cancellation mechanism of the supersymmetry variations, which are linear to $\psi$, 
for $\mathcal{N}=1$, $D=11$ supergravity can be sketched as follows.
\begin{alignat}{5}
  &\delta \mathcal{L}[eR] &&\sim [eR\bar{\epsilon}\psi], \notag
  \\
  &\delta \mathcal{L}[e\bar{\psi}\psi_{(2)}] &&\sim [eR\bar{\epsilon}\psi] \oplus
  &&[eF\bar{\epsilon}D\psi], \notag
  \\
  &\delta \mathcal{L}[eF^2] &&\sim &&&& [eDF\bar{\epsilon}\psi] \oplus
  &&[eF^2\bar{\epsilon}\psi],   
  \\
  &\delta \mathcal{L}[eF\bar{\psi}\psi] &&\sim &&[eF\bar{\epsilon}D\psi] \oplus
  &&[eDF\bar{\epsilon}\psi] \oplus && [eF^2\bar{\epsilon}\psi], \notag
  \\
  &\delta \mathcal{L}[AF^2] &&\sim &&&&&& [eF^2\bar{\epsilon}\psi]. \notag
\end{alignat}
We start from the Einstein-Hilbert term, $\mathcal{L}[eR]$.
First the variation of this term, $[eR\bar{\epsilon}\psi]$, is cancelled by the variation of
$\mathcal{L}[e\bar{\psi}\psi_{(2)}]$. Next to cancel the remaining variation 
$[eF\bar{\epsilon}D\psi]$, we need the term $\mathcal{L}[eF\bar{\psi}\psi]$ which is linear to 
the three form potential $A$. 
In a systematic way, we need to add $\mathcal{L}[eF^2] \propto A^2$ and 
$\mathcal{L}[AF^2] \propto A^3$ to cancel 
the variations $[eDF\bar{\epsilon}\psi]$ and $[eF^2\bar{\epsilon}\psi]$.
This is called Noether method and the coefficients of the action can be fixed completely.

The same strategy would also be applied to determine the contributions of
higher derivative corrections to the supergravity.
That is, we consider the supersymmetry transformations of $[eR^4]$ terms as
\begin{alignat}{5}
  &\delta \mathcal{L}[eR^4] &&\sim [eR^4\bar{\epsilon}\psi] \oplus \cdots, 
\end{alignat}
and fix the coefficients so as to cancel each term.
In the next section we will see this process in detail.

\section{Ansatz for Higher Derivative Terms and Variations under Supersymmetry}

From the calculations of string amplitudes in type IIA superstring theory, 
we know that corrections to type IIA supergravity arise from
quartic order of the Riemann tensor. Since these terms are trivially lifted to 
11 dimensions, it is natural to assume that corrections to 11 dimensional supergravity
also start from quartic order of the Riemann tensor. Then the ansatz for bosonic terms 
are written as
\begin{alignat}{3}
  &\mathcal{L}_B = \mathcal{L}[e R^4] + \mathcal{L}[\epsilon_{11} A R^4], 
  \\[0.2cm]
  &\mathcal{L}[e R^4] =
  + b^1_1 e R_{abcd}R_{abcd}R_{efgh}R_{efgh} 
  + b^1_2 e R_{abcd}R_{agfh}R_{becd}R_{efgh} \notag
  \\&\quad\qquad\quad
  + b^1_3 e R_{abcd}R_{abdh}R_{efcg}R_{efgh}
  + b^1_4 e R_{abcd}R_{aedg}R_{bcfh}R_{efgh} \notag
  \\&\quad\qquad\quad
  + b^1_5 e R_{abcd}R_{agdh}R_{bcef}R_{efgh}
  + b^1_6 e R_{abcd}R_{ahdf}R_{becg}R_{efgh} \notag
  \\&\quad\qquad\quad
  + b^1_7 e R_{abcf}R_{adgh}R_{bdce}R_{efgh}
  + b^1_8 e R_{abch}R_{adef}R_{bdcg}R_{efgh} \notag
  \\&\quad\qquad\quad
  + b^1_9 e R_{abch}R_{aedg}R_{bdcf}R_{efgh}
  + b^1_{10} e R_{abch}R_{aedf}R_{bcdg}R_{efgh} \notag
  \\&\quad\qquad\quad
  + b^1_{11} e R_{abch}R_{abdg}R_{cedf}R_{efgh}
  + b^1_{12} e R_{adgh}R_{afbc}R_{debc}R_{efgh} \notag
  \\&\quad\qquad\quad
  + b^1_{13} e R_{abch}R_{agde}R_{bdcf}R_{efgh}, \notag
  \\[0.1cm]
  &\mathcal{L}[\epsilon_{11} A R^4] =
  -\tfrac{1}{6} b^2_1 \epsilon_{11}^{\mu_1\cdots\mu_{11}} A_{\mu_1\mu_2\mu_3} 
  R_{ab\mu_4\mu_5}R_{ab\mu_6\mu_7}R_{cd\mu_8\mu_9}R_{cd\mu_{10}\mu_{11}} \notag
  \\&\qquad\qquad\quad\;\;
  -\tfrac{1}{6} b^2_2 \epsilon_{11}^{\mu_1\cdots\mu_{11}} A_{\mu_1\mu_2\mu_3} 
  R_{ab\mu_4\mu_5}R_{bc\mu_6\mu_7}R_{cd\mu_8\mu_9}R_{da\mu_{10}\mu_{11}}. \notag
\end{alignat}
There are important remarks on this ansatz.
First, as mentioned in the previous section, the Riemann tensor in $\mathcal{L}[e R^4]$ is
defined in terms of the supercovariant spin connection $\hat{\omega}(e,\psi)$ as
\begin{alignat}{3}
  R^{ab}{}_{\mu\nu}(\hat{\omega}) &= \partial_\mu \hat{\omega}_\nu{}^{ab} 
  - \partial_\nu \hat{\omega}_\mu{}^{ab}
  + \hat{\omega}_\mu{}^a{}_c \hat{\omega}_\nu{}^{cb}
  - \hat{\omega}_\nu{}^a{}_c \hat{\omega}_\mu{}^{cb}.
\end{alignat}
The Riemann tensor includes bilinear terms of the gravitino through the torsion part of 
the supercovariant spin connection, 
and the symmetry $R_{abcd}=R_{cdab}$ and the cyclicity $R_{a[bcd]}=0$ do not hold.
Thus the above thirteen terms contribute differently under the supersymmetry variation.
Next if we neglect the torsion part, because of the symmetry and the cyclicity of the Riemann tensor,
thirteen terms of $\mathcal{L}[eR^4]$ are redundant and classified by seven purely bosonic terms as
\begin{alignat}{3}
  \mathcal{L}[eR^4]_{\text{pure}} = &
  + e R_{abcd} R_{abcd} R_{efgh} R_{efgh} \times (b^1_1) &\qquad& A_1 \notag
  \\&
  + e R_{abcd} R_{abce} R_{dfgh} R_{efgh} \times (- \tfrac{1}{2}b^1_2 + b^1_3) &\qquad& A_2 \notag
  \\&
  + e R_{abcd} R_{abef} R_{cdgh} R_{efgh} \times 
  (- \tfrac{1}{8}b^1_4 - \tfrac{1}{4}b^1_5 + \tfrac{1}{8}b^1_6) &\qquad& A_3 \notag
  \\&
  + e R_{abcd} R_{aecg} R_{bfdh} R_{efgh} \times (- b^1_6) &\qquad& A_6 \label{eq:pboson}
  \\&
  + e R_{abce} R_{abdg} R_{cfdh} R_{efgh} \times (b^1_7 + b^1_8 + \tfrac{1}{2}b^1_9) &\qquad& A_5 \notag
  \\&
  + e R_{abce} R_{abdf} R_{cdgh} R_{efgh} \times 
  (\tfrac{1}{4}b^1_{10} - \tfrac{1}{2}b^1_{11} - b^1_{12}) &\qquad& A_4 \notag
  \\&
  + e R_{abce} R_{adcg} R_{bfdh} R_{efgh} \times (- b^1_9 - b^1_{13}). &\qquad& A_7 \notag
\end{alignat}
The notations $A_1, \cdots, A_7$ on the right hand side are used in ref.~\cite{RSW1}.
Finally we do not include terms which are proportional to the equations of motion for
the supergravity. These terms are removed by using the field redefinition ambiguities.

In order to cancel variations of these bosonic terms under the supersymmetric transformation,
we consider following fermionic terms.
\begin{alignat}{3}
  &\mathcal{L}_F = \mathcal{L}[e R^3 \bar{\psi} \psi_{(2)}] 
  + \mathcal{L}[e R^2 \bar{\psi}_{(2)} D \psi_{(2)}], 
  \\[0.2cm]
  &\mathcal{L}[e R^3 \bar{\psi} \psi_{(2)}] = 
  \sum_{n=1}^{92} f^1_n [e R^3 \bar{\psi} \psi_{(2)}]_n, \notag
  \\
  &\mathcal{L}[e R^2 \bar{\psi}_{(2)} D \psi_{(2)}] = 
  \sum_{n=1}^{25} f^2_n [e R^2 \bar{\psi}_{(2)} D \psi_{(2)}]_n. \notag
\end{alignat}
This ansatz is obtained by combining those in ref.~\cite{RSW2} and ref.~\cite{PVW}.
The explicit expressions of these terms will be given in ref.~\cite{HO}.

Now let us consider the variation of the Lagrangian under the local supersymmetry.
For bosonic terms we use the variation of the Riemann tensor of the form
\begin{alignat}{3}
  \delta R_{abcd} &= e^\mu{}_c e^\nu{}_d \delta R_{ab\mu\nu}
  - R_{abd\mu} \delta e^\mu{}_c + R_{abc\nu} \delta e^\nu{}_d \notag
  \\
  &= D_c \delta \hat{\omega}_{dab} - D_d \delta \hat{\omega}_{cab} 
  + R_{abde} \bar{\epsilon} \gamma^e \psi_c - R_{abce} \bar{\epsilon} \gamma^e \psi_d 
  + \mathcal{O}(\psi^3).
\end{alignat}
The covariant derivative acts on all local Lorentz indices.
Since we focus on the cancellation of terms which linearly depend on the gravitino,
we only consider the bosonic part of the supercovariant spin connection and 
the Riemann tensor on the right hand side.
Let us vary $b^1_1$ term of $\mathcal{L}[eR^4]$ as an example.
\begin{alignat}{3}
  \delta(b^1_1 e R_{abcd}R_{abcd}R_{efgh}R_{efgh}) &\sim
  + b^1_1 e R_{abcd}R_{abcd}R_{efgh}R_{efgh} \bar{\epsilon} \gamma^z \psi_z \notag
  \\
  &\quad -8 b^1_1 e R_{iefg}R_{zefg}R_{abcd}R_{abcd} \bar{\epsilon} \gamma^i \psi_z
  \\
  &\quad -32 b^1_1 e R_{ijka}R_{ebcd}D_eR_{abcd} \bar{\epsilon} \gamma^k \psi^{ij}. \notag
\end{alignat}
To obtain this expression we used partial integral and Bianchi identity $D_{[e}R_{ab]cd}=0$.
We also dropped the terms which are proportional to the equations of motion, since
these terms can be cancelled by modifying the supersymmetry transformations.
Variations of fermionic parts are obtained in a similar way. 
Because there are more than one hundred terms, we used a computer programming to 
calculate the variations.

Since we are interested in the cancellation of the terms which do not include the three form potential,
for the terms in $\mathcal{L}[eAR^4]$ we should take the variation
for the three form potential.
Then the variation of the Lagrangian $\mathcal{L} = \mathcal{L}_{B} + \mathcal{L}_{F}$
under the supersymmetry transformation is sketched as
\begin{alignat}{5}
  &\delta \mathcal{L}[eR^4] &&\sim [eR^4\bar{\epsilon}\psi] \oplus
  &&[eR^2DR\bar{\epsilon}\psi_{(2)}], \notag
  \\
  &\delta \mathcal{L}[\epsilon_{11}AR^4] &&\sim [eR^4\bar{\epsilon}\psi], \notag
  \\
  &\delta \mathcal{L}[eR^3\bar{\psi}\psi_{(2)}] &&\sim [eR^4\bar{\epsilon}\psi] \oplus
  &&[eR^2DR\bar{\epsilon}\psi_{(2)}] \oplus &&[eR^3\bar{\epsilon}D\psi_{(2)}], \label{eq:var}
  \\
  &\delta \mathcal{L}[e R^2 \bar{\psi}_{(2)} D \psi_{(2)}] 
  &&\sim &&[eR^2DR\bar{\epsilon}\psi_{(2)}] \oplus &&[eR^3\bar{\epsilon}D\psi_{(2)}], \notag
  \\
  &0 &&\sim [eR^4\bar{\epsilon}\psi] \oplus &&&& [eR^3\bar{\epsilon}D\psi_{(2)}]. \notag
\end{alignat}
There are 116 terms for $[eR^4\bar{\epsilon}\psi]$, 88 terms for 
$[eR^2DR\bar{\epsilon}\psi_{(2)}]$ and 60 terms for $[eR^3\bar{\epsilon}D\psi_{(2)}]$.
Note, however, that the first terms $[eR^4\bar{\epsilon}\psi]$ and 
the third terms $[eR^3\bar{\epsilon}D\psi_{(2)}]$ in the variations 
are not independent because of the identity:
\begin{alignat}{3}
  D_{[e} \psi_{cd]} &= \tfrac{1}{4} \gamma^{ab} \psi_{[e} R_{cd] ab}.
\end{alignat}
In fact, by using the computer program we find that there are 20 identities between the first terms
and the third terms. 
\begin{alignat}{3}
  0 &= \sum_{i=n}^{20} i_n ([R^4 \bar{\epsilon} \psi]_n
  + [R^3 \bar{\epsilon} \mathcal{D} \psi_{(2)}]_n).
\end{alignat}
The last line of eq.~(\ref{eq:var}) represents these identities.

Now under the local supersymmetry transformation, we obtain $264$ independent terms.
And the coefficients of $264$ terms are finally given by linear combinations of 
$b^1_n$, $b^2_n$, $f^1_n$, $f^2_n$ and $i_n$.
Therefore by requiring the supersymmetry, we obtain 264 linear equations among 
152 variables of $b^1_n$, $b^2_n$, $f^1_n$, $f^2_n$ and $i_n$.

\section{Results}

We solved the linear equations among $b^1_n$, $b^2_n$, $f^1_n$, $f^2_n$ and $i_n$
by using computer programming and found only two parameters which relate bosonic terms and fermionic 
terms\footnote{We used mathematica for our programming, which employed the package by U. Gran\cite{Gr}.}.
Here we only show results for the coefficients $b^1_n$ and $b^2_n$.

The first solution is given by
\begin{alignat}{3}
  (b^1_1, b^1_2, b^1_4, b^1_6, b^1_7, b^1_9, b^2_1, b^2_2) = 
  a(1,32,-32,-16,-16,-32,-\tfrac{1}{4},1).
\end{alignat}
This relates terms in $\mathcal{L}[eR^4]$ and $\mathcal{L}[\epsilon_{11}AR^4]$.
The purely bosonic coefficients are obtained by using the eq.(\ref{eq:pboson}).
\begin{alignat}{3}
  \mathcal{L}_{B}^{\text{pure}} &=
  a \big( + e R_{abcd} R_{abcd} R_{efgh} R_{efgh} 
  - 16 e R_{abcd} R_{abce} R_{dfgh} R_{efgh} \notag
  \\&\qquad\;
  + 2 e R_{abcd} R_{abef} R_{cdgh} R_{efgh} 
  + 16 e R_{abcd} R_{aecg} R_{bfdh} R_{efgh} \notag
  \\&\qquad\;
  - 32 e R_{abce} R_{abdg} R_{cfdh} R_{efgh} 
  + 32 e R_{abce} R_{adcg} R_{bfdh} R_{efgh} \notag
  \\&\qquad\;
  + \tfrac{1}{24} \epsilon_{11}^{\mu_1\cdots\mu_{11}} A_{\mu_1\mu_2\mu_3} 
  R_{ab\mu_4\mu_5}R_{ab\mu_6\mu_7}R_{cd\mu_8\mu_9}R_{cd\mu_{10}\mu_{11}} \notag
  \\&\qquad\;
  - \tfrac{1}{6} \epsilon_{11}^{\mu_1\cdots\mu_{11}} A_{\mu_1\mu_2\mu_3} 
  R_{ab\mu_4\mu_5}R_{bc\mu_6\mu_7}R_{cd\mu_8\mu_9}R_{da\mu_{10}\mu_{11}} \big) \notag
  \\[0.1cm]
  &= \tfrac{1}{12}a \big( t_8 t_8 e R^4 - \tfrac{1}{12} \epsilon_{11} t_8 A R^4 \big). \label{eq:a}
\end{alignat}
This precisely matches with the first term in eq.~(\ref{eq:sinv}).
The explicit expression for the fermionic part is given in ref. \cite{PVW}.
We also checked the fermionic part of our calculation coincide with their result.

The second solution is expressed as
\begin{alignat}{3}
  (b^1_7,b^1_9,b^1_{10},b^1_{11},b^1_{12}) = 
  b(\tfrac{1}{2},-1,\tfrac{1}{5},\tfrac{1}{5},\tfrac{1}{5}).
\end{alignat}
Again the purely bosonic coefficients are obtained by using the eq.(\ref{eq:pboson}).
\begin{alignat}{3}
  \mathcal{L}_{B}^{\text{pure}} &= 
  b \big(- \tfrac{1}{4} e R_{abce} R_{abdf} R_{cdgh} R_{efgh} 
  + e R_{abce} R_{adcg} R_{bfdh} R_{efgh} \big) \notag
  \\
  &= \tfrac{1}{24 \times 32}b \big( t_8 t_8 e R^4 
  + \tfrac{1}{4!} \epsilon_{11} \epsilon_{11} e R^4 \big). \label{eq:b}
\end{alignat}
This completely matches with the second term in eq.~(\ref{eq:sinv}).
Therefore we could derive the bosonic terms of two superinvariant completely by employing
$\mathcal{N}=1$ supersymmetry in eleven dimensions.

We also found 13 parameters which only relate the coefficients of fermionic terms\cite{HO}. 
These parameters will be fixed by considering the cancellation of variations which 
depend on the three form potential.

\section{Conclusions and Discussions}

By applying Noether method to $\mathcal{N}=1$ local supersymmetry in eleven dimensions,
we obtained two candidates of higher derivative corrections to 
$\mathcal{N}=1$, $D=11$ supergravity. The bosonic parts of these two completely match with 
the results obtained by type IIA string perturbative calculations. 
Thus we can fix the bosonic parts of superinvariants only by employing the local supersymmetry.

On the other hand, we still have 13 parameters which only relate the coefficients of the fermionic terms.
By choosing these parameters appropriately we can realize the result obtained in ref.~\cite{PVW}.
In order to restrict these parameters we should proceed to the cancellation of the variations which 
include the three form potential. This is our future work. It is interesting to relate our results
to the superfield method \cite{CGNN,HSS,Ra}.
Applications to black hole physics and cosmology are also important directions\cite{My,MO}.

\vspace*{0.5cm}
\noindent
{\bf [Note added]}

In refs.~\cite{RSW1,RSW2}, Roo et al. constructed two superinvariants in ten dimensions 
which are related to the Green Schwarz anomaly cancellation terms. 
One of them includes terms like $[HR^2DR]$,
where $H$ represents the field strength of the NS-NS $B$ field.
On the other hand, we could construct one superinvariant in eleven dimensions 
which is related to the Green Schwarz anomaly cancellation terms.
This difference occurs because we do not include terms like $[FR^2DR]$ in our ansatz.
In fact these terms should not exist since we cannot find fermionic counter terms to
cancel the variations. Furthermore these terms violate the parity invariance of 
M-theory\footnote{We would like to thank Pierre Vanhove for communicating with this issue.}.

\section*{Acknowledgements}

We would like to thank Gary Horowitz for many useful discussions.
This research was supported in part by the Japan Society for the Promotion of Science. 
This research was also supported in part by the National Science Foundation under Grant No. PHY99-07949.


\begin{thebibliography}{99}


\bibitem{GW} 
D. J. Gross and E. Witten,
Nucl. Phys. {\bf B277} (1986) 1.

\bibitem{GSl} 
D. J. Gross and J. H. Sloan,
Nucl. Phys. {\bf B291} (1987) 41.

\bibitem{GVZ} 
M. T. Grisaru, A. E. van de Ven and D. Zanon,
Phys. Lett. {\bf B173} (1986) 423.

\bibitem{KP}
E. Kiritsis and B. Pioline,
Nucl. Phys. {\bf B508} (1997) 509,
[arXiv:hep-th/9707018].

\bibitem{VW}
C. Vafa and E. Witten,
Nucl. Phys. {\bf B447} (1995) 261-270,
[arXiv:hep-th/9505053]. 

\bibitem{DLM}
M. J. Duff, J. T. Liu and R. Minasian,
Nucl. Phys. {\bf B452} (1995) 261-282,
[arXiv:hep-th/9506126]. 

\bibitem{GS} 
M. B. Green and J. H. Schwarz, 
Phys. Lett. {\bf B149} (1984) 117.

\bibitem{Ts} 
A. A. Tseytlin, 
Nucl. Phys. {\bf B584} (2000) 233, 
[arXiv:hep-th/0005072]. 

\bibitem{RSW1}
M. de Roo, H. Suelmann and A. Wiedemann,
Phys. Lett. {\bf B280} (1992) 39. 

\bibitem{RSW2}
M. de Roo, H. Suelmann and A. Wiedemann,
Nucl. Phys. {\bf B 405}(1993) 326,
[arXiv:hep-th/9210099]. 

\bibitem{Su}
H. Suelmann, 
PhD Thesis, Groningen University, 1994.

\bibitem{PVW}
K. Peeters, P. Vanhove and A. Westerberg,
Class. Quant. Grav. {\bf 18} (2001) 843,
[arXiv:hep-th/0010167]. 

\bibitem{PVW2}
K. Peeters, P. Vanhove and A. Westerberg,
Class. Quant. Grav. {\bf 19} (2002) 2699,
[arXiv:hep-th/0112157]. 

\bibitem{CJS} 
E. Cremmer, B. Julia and J. Scherk,
Phys. Lett. {\bf B76} (1978) 409.

\bibitem{HO}
Y. Hyakutake and S. Ogushi, in preparation.

\bibitem{Gr}
U. Gran,
[arXiv:hep-th/0105086]. 

\bibitem{CGNN}
M. Cederwall, U. Gran, M. Nielsen and B. E. Nilsson,
JHEP {\bf 0010} (2000) 041,
[arXiv:hep-th/0007035]. 

\bibitem{HSS}
S. de Haro, A. Sinkovics and K. Skenderis, 
Phys. Rev {\bf D67} (2003) 084010,
[arXiv:hep-th/0210080]. 
 
\bibitem{Ra} 
A. Rajaraman,
[arXiv:hep-th/0505155]. 

\bibitem{My}
R. Myers,
Nucl. Phys. {\bf B289} (1987) 701.

\bibitem{MO}
Kei-ichi. Maeda and N. Ohta,
Phys. Lett. {\bf B597} (2004) 400,
[arXiv:hep-th/0405205]. 




\end{thebibliography}
\end{document}